\documentclass[letterpaper, 10 pt, conference]{ieeeconf}  % Comment this line out
\IEEEoverridecommandlockouts                              % This command is only
\usepackage{amsmath}
\usepackage{subfigure}
\usepackage{graphicx}
\usepackage{url}
\usepackage{algorithm}
\usepackage{algorithmic}
\usepackage{amssymb}
\def\xx{{\boldsymbol{x}}}

\def\N{{\mathcal{N}}}  
\def\X{{\boldsymbol{X}}}  
 
 \def\RR{{\bf R}}
\def\AA{{\boldsymbol{A}}}

\def\zz{{\boldsymbol{z}}} 
 \def\yy{{\boldsymbol{y}}}
\def\bb{{\boldsymbol{b}}}

\def\vv{{\boldsymbol{v}}}  
\def\ww{{\boldsymbol{w}}}  
\def\PP{{\boldsymbol{P}}}  
  
\def\Q{{\boldsymbol{Q}}}  

\def\subjto{{\mbox{subj. to}}}

\newcommand{\Ce}{\mathbb{C}}

\RequirePackage{xspace} \RequirePackage{amsmath, amssymb,color}

\DeclareMathOperator*{\Tr}{Tr}

 \def\xx{{\boldsymbol{x}}}

 \def\zz{{\boldsymbol{z}}} 
 \def\yy{{\boldsymbol{y}}}
\def\bb{{\boldsymbol{b}}}  
\def\RR{{\boldsymbol{R}}}  
  
\def\ss{{\boldsymbol{s}}}

\def\subjto{{\mbox{subj. to}}}

\renewcommand{\Re}{{\mathbb{R}}}
\newcommand{\T}{\mathsf{T}}
\newcommand{\eg}{\textit{e.g.,~}}
\newcommand{\ie}{\textit{i.e.,~}}

\def\I{{\boldsymbol{I}}}

\title{\LARGE \bf
Nonlinear Compressive Particle Filtering 
}

\author{Henrik Ohlsson,  Michel Verhaegen and S. Shankar Sastry% <-this % stops a space
\thanks{The authors gratefully acknowledge support by the Swedish Research
  Council in the Linnaeus center CADICS, the European Research Council
   under the advanced grant LEARN, contract 267381,  a postdoctoral grant from the Sweden-America
   Foundation, donated by ASEA's Fellowship Fund,  a postdoctoral
   grant from the Swedish Research Council and the Dutch National Science Foundation STW.}% <-this % stops a space
\thanks{Ohlsson, and Sastry are with the Department of Electrical Engineering and Computer  Sciences, University of California, Berkeley, CA, USA. Ohlsson is also with the
Division of Automatic Control, Department of Electrical Engineering,
Link\"oping University, Sweden. Verhaegen is with the Delft Center for Systems and Control, Delft University, Delft 2628CD, The Netherlands.
Corresponding author: Henrik Ohlsson, Cory Hall, University of California, Berkeley, CA 94720. Email: 
{\tt\small  ohlsson@eecs.berkeley.edu.}}}%

\begin{document}

\maketitle
\thispagestyle{empty}
\pagestyle{empty}

%%%%%%%%%%%%%%%%%%%%%%%%%%%%%%%%%%%%%%%%%%%%%%%%%%%%%%%%%%%%%%%%%%%%%%%%%%%%%%%%
\begin{abstract}
Many systems for which compressive sensing is used today are dynamical. The
common approach is to neglect the dynamics and see the problem as a
sequence of
independent problems. This approach has two disadvantages. Firstly,
the temporal dependency in the state could be
used to improve the accuracy of the state estimates. Secondly,  having
 an estimate for the state and its support could be used to reduce the computational
 load of the subsequent step.

In the linear Gaussian setting,  compressive sensing was recently combined with the Kalman filter to
 mitigate above disadvantages. In the nonlinear dynamical case, compressive
 sensing can not be used and, if the state dimension is high, the
 particle filter would perform poorly. In this paper we combine one of
 the most novel developments in compressive sensing, nonlinear
 compressive sensing, with the particle filter. We show that the
 marriage of the two is essential and that neither the particle filter
 or nonlinear compressive sensing alone gives a satisfying solution.

\end{abstract}

%%%%%%%%%%%%%%%%%%%%%%%%%%%%%%%%%%%%%%%%%%%%%%%%%%%%%%%%%%%%%%%%%%%%%%%%%%%%%%%%
\section{Introduction}
\textit{Compressive sensing} (CS) \cite{Candes:06,Donoho:06}  has gained considerable attention the last
decade. Its remarkable property of exactly estimating a
linear system
from far fewer measurements than what previously thought possible has
inspired to many new interesting developments and applications.

The holy grail of compressive sensing is the ability to solve
optimization problems of the form
\begin{equation}
\min_\xx \|\xx\|_0 \quad \subjto \quad \yy=\AA\xx.
\end{equation}
$\xx\in \Re^n$ is here the sought parameter vector, $\AA\in \Re^{N
  \times n}$ a sensing matrix which is generally assumed known,  and
$\yy \in \Re^{N}$ a given vector of stacked measurements. For a
detailed review of the literature, the reader is referred to several
recent publications, such as \cite{bruckstein:09} and \cite{Eldar:2012}.

One of
the newest developments in compressive sensing is \textit{nonlinear
compressive sensing} (NLCS) \cite{Beck:2012,BlumensathT:2012,ohlsson:13}. In NLCS  the measurement equation is not
required to be linear. This is particularly interesting in
applications where physical limitations make the measurement equation
nonlinear. In \eg X-ray crystallography and many applications in
optics, only intensities are observed. Since $\xx$ in these
applications often is a complex valued vector, the phase information is lost when
measuring. The measurement equation then
takes the form 
\begin{equation}\yy=|\AA\xx|^2, \quad \yy\in \Re^N,\;\AA \in \Ce^{N\times
    n},\;\xx\in\Ce^{n},  \end{equation} where the absolute value and
the square are acting elementwise on the complex vector $\AA \xx$.
Other applications with this type of nonlinear measurement equation show up in
\eg diffraction imaging \cite{Bunk:07},
astronomy \cite{Dainty:87,Fienup:93},
optics \cite{WaltherA1963},
x-ray tomography \cite{Dierolf:10},
microscopy \cite{miao:08,Antonello:12,Szameit:12},
and quantum mechanics \cite{Corbett:06} and the problem is often
referred to as the \textit{phase
  retrieval problem}.

Many systems that are of interest in compressive sensing are dynamical
systems. The temporal relation between $\xx$s are often neglected and
compressive sensing applied repeatedly as if temporally consecutive
$\xx$s would be independent.  It is rather obvious that taking into account
the temporal relation between two consecutive $\xx$s might
improve estimation accuracy over treating  two consecutive $\xx$s as independent
vectors. It might also be that an estimate of $\xx$ at time $t$ can
be used to reduce the
computational load  at time $t+1$.

% Many systems of interest for engineers  are dynamic. A very general  model for dynamical systems
% is given by
% \begin{subequations}
% \begin{align}
% \xx(t+1) = &g(\xx(t),t)+v(t)\\
% \yy(t)=&h(\xx(t),t)+w(t)
% \end{align}\end{subequations}
% The goal of filtering, is to estimate $\xx(t),\,t=1,\dots, T$ from
% the measurements $\{\yy(t)\}_{t=1}^T,$ and the functions
% $g(\cdot,t),\, h(\cdot,t),\,t=1,\dots,T.$ The noise distributions for
% the process noise $w$ and the measurement noise $v$ are also assumed
% known. 

% Extend compress sensing and nonlinear compressive sensing to
% filtering could have many advantages. The main motivation might be
% the improved accuracy. It is rather obvious that taking into account
% that two consecutive $\xx$s are related by some process model might
% improve accuracy over treating  two consecutive $\xx$s as independent
% vectors. It is also well known that compressive sensing can provide
% very accurate estimates under the assumption that the sought quantity
% is sparse. For a sparse $\xx$, a compressive sensing framework that
% takes into account the temporal relation between $\xx$s is therefore
% very interesting. But a often forgotten motivation is that the
% computational complexity of filters grows very fast with the dimension
% of the state. 

% If we assume that $\xx(t),\,t=1,\dots, T$ is a sparse sequence of
% vectors. Nonlinear compressive sensing could be applied repeatedly on
% the measurement equation
% \begin{equation} \yy(t)=h(\xx(t),t)+w(t).\end{equation}
% However, it is unclear how the dynamics should be taken into account. 

In this paper we propose to combine nonlinear compressive sensing with
 particle filtering to estimate a sequence of sparse temporally dependent
states. We will assume that the state support is unknown and estimate the
state $\xx$ and its support from measurements $\yy$. We will show that this approach outperforms the alternative of
neglecting the temporal relation between consecutive $\xx$s. We also
show that the problem can not be solved by a particle filter alone
and hence that the marriage of the two is essential.

% In this paper we 
% treat and analyze the problem of ``sparse estimation framework'' with linear
% dynamics and nonlinear measurement equations in a particle filtering
% framework. Following our recent contribution to Quadratic Basic
% Pursuit, we assume the nonlinear output equation to be a general
% quadratic form \cite{Ohlsson13}. The
% organization fo this paper is ...

\section{Background}

Nonlinear state estimation is a system theoretic problem of great
scientific importance in many fields and application domains. In
general the estimation problems are an extension of static nonlinear
parameter estimation problems where the unknown parameter vector obeys
a given dynamic propagation law. Such propagation laws can be given in
either continuous or discrete time by a nonlinear differential or
difference equation. A state estimation problem is characterized via
three classes of information: (a) the dynamic propagation law, (b) the
measurement equation characterizing the relationship between the
measured quantity and the unknown state vector and (c) the noise
quantities characterizing (one of) the driving forces of the dynamic
propagation law and the measurement errors made. In this paper we do
not consider deterministic (known) driving forces for the dynamic
propagation laws. 

Despite the major interest, a generic solution to the nonlinear state reconstruction
problems for nonlinear models for (a) and (b) and arbitrary noise
statistics is computational untractable. Therefore succesful solutions
consists in specializing one or more of the information classes in the
definition of nonlinear state estimation problems to ``restrictive''
circumstances and/or reciding to ``approximate solutions''. One such
famous family of algorithms are the so-called Extended Kalman filters
where smooth nonlinear functions in the description (a) and (b) are
assumed while the noise statistics are approximated by Gaussian 
probability density functions (see for instance \cite{Sorenson85}). Various improvements on these
restrictions and/or assumptions were made since then. Here we mention
the unscented Kalman
filter (UKF) \cite{Julier95}  and the derivative free nonlinear Kalman
filter \cite{Rigatos2011}, set-membership state estimation \cite{Scholte2003}, Ellipsoid state estimation
\cite{Chernousko2005} and the particle filter \cite{gordon:93,Vaswani08a}.

An important specialization of the state reconstruction problem is for
the case the unknown state vector is sparse. Thereby calling for a
dynamic extension to the succesfull research line of reconstruction of a large unknown
parameter vector in linear least squares problems from a limited
number of measurements by imposing  sparsity constraints, as in
compressed sensing \cite{Candes:06,Donoho:06}.   Such a dynamic extension for
linear dynamics and linear measurement equation has recently been
studied in \cite{Vaswani08b,wang:09,Carmi10,das:13}. In these recent
approaches, we recognize two families of approaches to perform
``sparse state reconstruction''. The first, like \cite{Carmi10}
integrates the $\ell_1$ norm constraints as a pseudo measurement in a
classical minimum variance or $2$ norm (extended) Kalman filtering
framework. Even for the case the dynamic propagation and measurement
equation are linear, the $2$-norm approximation requires iterations
inheriting all advantages {\em and} disadvantages of the extended
Kalman filtering. 
The second class of filtering methods try in
addition to the reconstruction of the state, to estimate its support
as well.  The sparsity estimation was dealt with in 
in \cite{wang:09} 
with a multiple of particle filters, whereby the large dimensional output
vectors was mapped onto a lower dimensional space using classical ideas
of compressive sensing. 
In \cite{Vaswani08b} a
combination of a classical generalized likelihood ratio testing for
expanding (or decreasing) the unknown support set of the state vector
is combined with classical reduced Kalman filtering.
Later on the same author and collaborators in \cite{das:13} have adapted the
PF with an importance sampling step for predicting the possibly
``abrupt'' changes in the support of the state. That support
prediction is based on classical $\ell_1$ regularization compressed
sensing solutions for static  linear measurement equations.
 In this paper a
generalization to particle filters for  nonlinear measurement equations
is presented.

% \section{Notation and Assumptions}

% We will in general use bold face to denote vectors and matrices and normal font
% for scalars. We denote
% the transpose of a real vector by  $\xx^\T$.  $\X_{i,j}$ is used to denote the
% $(i,j)$th element, $\X_{i,:}$ the $i$th row and  $\X_{:,j}$ the $j$th
% column of a matrix $\X$, respectively. We will use the notation
% $\X_{i_1:i_2,j_1:j_2}$ to denote a submatrix constructed from
% rows $i_1$ to $i_2$ and columns  $j_1$ to $j_2$ of $\X$. Given two matrices $\X$ and $\Y$,
% we use the following fact that their product in the trace function commutes, namely, $\Tr(\X \Y) = \Tr(\Y \X)$, under the assumption
% that the dimensions match. $\| \cdot \|_0$ counts the number of nonzero elements in a vector or matrix; similarly, \linebreak $\| \cdot\|_1$ denotes the element-wise $\ell_1$-norm of a vector or matrix, \ie, the sum of the magnitudes of the elements; whereas $\| \cdot \|$ represents the $\ell_2$-norm for vectors and the spectral norm for matrices.
%  We will assume that $f_i(\cdot),i=1,\dots,N$ are analytic functions. 

\section{Setup}

%\subsection{Dynamical Model}
We will assume  that the state $\xx$ follows a nonlinear process model
with additive process noise
and that the support (\ie the set of indices corresponding to nonzero
elements) of $\xx$ is slowly varying. The state $\xx$  will have
dimension $n$ and the measurements $\yy$ dimension $N$.    We will restrict ourself to
quadratic measurement equations. We do this to ease the notation and also
because we believe that it is an important special case that provides
a sufficiently
good approximation of the nonlinearity in many applications of interest. As
shown in \cite{ohlsson:13b}, the extension to handle more general
nonlinearities is rather trivial. We will for simplicity assume that
the measurement  noise $\ww$ is additive and Gaussian. The process
noise $\vv$ will be assumed additive and with a \textit{probability density
function} (pdf) given by $p_{\vv}(\cdot)$. $g:\Re^n \rightarrow \Re^n$ is the process
function and $h:\Re^n \rightarrow \Re^N$ the measurement function with
the $i$th element
\begin{equation*}
\big (h(\xx)  \big )(i) \hspace{-0.1cm}= \hspace{-0.1cm} a_i+\bb_i^\T \xx +
\xx^\T  \Q_i \xx,\, a_i\in \Re, \bb_i \in \Re^n,\Q_i\in
\Re^{n\times n}.
\end{equation*}
 $diag(\cdot)$ produces a quadratic
matrix with its argument on the diagonal. In particular, if we  let $\ss(t)$
be a binary row vector of length $n$, the dynamical nonlinear model
takes the form: 
\begin{subequations}\label{eq:dyn}
\begin{align}\label{eq:dyneq}
\xx(t+1) = &diag(\ss(t)) \left (g(\xx(t),t)+\vv(t) \right),\\
\yy(t)=&h(\xx(t))+\ww(t),\\
\ww(t) \sim & \N(0,\RR),\, \vv(t)\sim p_{\vv},\, \xx(t) \in \Re^n,\,\yy(t) \in \Re^N.
%\\ h(\xx(t) ,t)=& a(t)+\bb(t)^\T \xx(t) +
%\xx^\T (t) \Q(t) \xx(t)
%\\  %\vv(t) \sim&\N(0,\QQ(t)),  \quad 
%\ww(t) \sim&
%\mathcal{U}(-\epsilon,\epsilon).
\end{align}\end{subequations}
For each of the  elements in $\ss(t)$, the binary value stays the same
with probability $\alpha$ and the element is changed with probability $1-
\alpha$ (basically a Markov chain).  It is assumed that $\alpha$ is close to
1.  The indices of the nonzero elements of $\ss(t)$ make up the
support.  We will make the assumption that the number of elements in
the support is rather
small in comparison to the number of measurements $N$ and the state
dimension $n$. 
%Note that this means that most states will be zero and only a few
%nonzero.

We will represent the distribution for the state $\xx$ using particle
approximations. If we denote the $i$th particle at time $t$ by
$\xx^{(i)}(t)$ and its corresponding weight by $w^{(i)}(t)$, the
particle approximation using $M$ particles of the pdf of $\xx(t)$
given $\ss(t)$ and measurements up to time $t$ is 
\begin{align}\nonumber
p(\xx(t)|\{\yy(s&)\}_{s=1}^{t} ,\ss(t)) \\ \label{eq:filterpdf} \approx& \sum_{i=1}^M
w^{(i)} (t) \delta \left (\xx(t) -diag(\ss(t))\xx^{(i)}(t) \right).
  \end{align}
$\delta(\cdot)$ is here used to denote  the delta Dirac function. Similarly, we will let the particle approximation using $M$ particles
of the pdf of the predicted state at time $t+1$, $\xx(t+1|t)$, 
given $\ss(t)$ and measurements up to time $t$, be given by 
\begin{align}\nonumber
&p(\xx(t+1|t)|\{\yy(s)\}_{s=1}^{t},\ss(t)) \\ \label{eq:partapp} & \approx  \sum_{i=1}^M w^{(i)}
(t+1|t) \delta \left(\xx(t+1|t) -diag(\ss(t))\xx^{(i)}(t+1|t) \right).
  \end{align}
Note that $\xx(t+1|t)$ is used to denote the predicted state at time
$t+1$ given measurements up  to time $t$. We accordingly let $w^{(i)}
(t+1|t)$ denote the predicted weight at time $t$ of particle $i$  and  $\xx^{(i)}(t+1|t)$
the predicted $i$th particle at time $t+1$, both using measurements up to time $t$.

\section{Nonlinear Compressive Particle Filtering}

In this paper we 
 treat and analyze the problem of sparse estimation with nonlinear
 dynamics and quadratic measurement equations by combining
 ideas from particle filtering and nonlinear compressive sensing, and
 in particular \textit{quadratic basis
 pursuit} (QBP) \cite{ohlsson:13}.  We believe that a second order Taylor
expansion of the measurement function often is sufficient and
therefore restrict the discussion to quadratic measurement functions.

We will describe the \textit{nonlinear compressive particle filter} (NCPF) in
detail in the following subsections, however, we start by giving a
brief  outline of the algorithm:
\begin{itemize}
\item[0)] {\bf Initialization: }Given an estimate for $\ss(0)$ and
  the number of particles to be used $M$.  Generate
  $\{\zz^{(i)}\}_{i=1}^M \in \Re^n$ by sampling from some given
  initial distribution $p_0$. Initialize the particles
  $\{\xx^{(i)} (1|0)\}_{i=1}^M \in \Re^n$ by $\xx^{(i)}(1|0)=diag(\ss(0))\zz^{(i)},\,i=1,\dots,M$.
Initialize the weights by   $\{w^{(i)} (1|0)\}_{i=1}^M =1/M$. Form a
particle approximation of the predictive distribution $p(\xx(1|0) |
\ss(0))$ by \eqref{eq:partapp}.
Set the design parameter $\Delta t$, $\lambda$ and $\epsilon$. When
  $\yy(1)$ becomes available,  set $t=1$ and proceed to  step~1.
\item[1)]{\bf Propagate the Predictive Distribution
 One Step: } Given $\yy(t)$ and a particle approximation of the predictive distribution at time $t$,
$p(\xx(t|t-1)|\{\yy(s)\}_{s=1}^{t-1}, \ss(t-1))$, compute a particle
approximation  of the predictive distribution
at time $t=t+1$, 
$p(\xx(t+1|t)|\{\yy(s)\}_{s=1}^{t},\ss(t-1))$, using $\ss(t-1)$. 
Proceed to  step 2.
%Run the particle filter on the part of $\xx$ that is nonzero. 
\item[2)] {\bf Add Elements to the Support: } If the likelihood for
  the observations $\yy(t)$, $p(\yy(t) |\hat \xx(t) ) $, where $\hat \xx$
  is an approximation of the expected value of $\xx(t)$, falls below some
  threshold, apply ideas from NLCS to detect what new elements of $\xx$ are likely
  to be nonzero and add their indices to the
  support by updating $\ss(t-1)$. Proceed to  step~3.
\item[3)]{\bf Remove Elements from the Support: } If the mean of an element of  $\xx$ gets within some $\epsilon$-bound from
  zero and remains within this bound for $\Delta t$  consecutive time
  steps, exclude that element from the set of nonzero elements by
  setting corresponding element in  $\ss(t-1)$ to zero. 
\item[4)]{\bf Update the Predictive Distribution: }If the
  support has changed, set $\ss(t-\Delta t-1)=\ss(t-1)$, compute
  $p(\xx(t-\Delta t|t-\Delta t-1 )|\{\yy(s)\}_{s=1}^{t-\Delta t-1}, \ss(t-\Delta t-1))$, set $t=t-\Delta t$, and return to step
  1. Otherwise, set $\ss(t-1)=\ss(t)$, $p(\xx(t+1|t)|\{\yy(s)\}_{s=1}^{t}, \ss(t))=p(\xx(t+1|t)|\{\yy(s)\}_{s=1}^{t}, \ss(t))$, $t=t+1$ and return to step 1. 
\end{itemize}

% \begin{algorithm}
%   \caption{Counting mismatches between two packed \DNA strings
%     \label{alg:packed-dna-hamming}}
%   \begin{algorithmic}[1]
%     \Require{$x$ and $y$ are packed \DNA strings of equal length $n$}
%     \Statex
%     \Function{Distance}{$x, y$}
%       \Let{$z$}{$x \oplus y$} \Comment{$\oplus$: bitwise exclusive-or}
%       \Let{$\delta$}{$0$}
%       \For{$i \gets 1 \textrm{ to } n$}
%         \If{$z_i \neq 0$}
%           \Let{$\delta$}{$\delta + 1$}
%         \EndIf
%       \EndFor
%       \State \Return{$\delta$}
%     \EndFunction
%   \end{algorithmic}
% \end{algorithm}

We will in the following five subsections describe each of these steps in detail.

\subsection{Step 0: Initialization}
We will assume that at time $0$ we are given an estimate for the
support $\ss(0)$. This could for example be computed by seeking the sparsest
solution satisfying the measurement equation at time $t=1$.  
%\begin{equation}\label{eq:quad}
%\min_{\xx} \|\xx\|_0 + \lambda \| y(0) - h(\xx,0) \|_{\RR(0)}^2.
%\end{equation} 
Since the measurement function $h$ is assumed to be a quadratic function,
QBP \cite{ohlsson:13} provides a suitable method for finding an estimate
of the initial  support. %Generate 
%$\{\zz^{(i)}\}_{i=1}^M$
%from  $p_0$ and 
Initialize the particles
$\{\xx^{(i)}(0)\}_{i=1}^M$ by  sampling $\zz^{(i)},\,i=1,\dots,M,$ from
$p_0$ and then set $\xx^{(i)}(1|0) =diag(\ss(0))\zz^{(i)} ,\,i=1,\dots,M$. %(\xx(t)|s()) %setting  $\xx^{(i)}(0)=diag(\ss(0))
%\zz^{(i)}, i=1,\dots, M$. 
%We will use superindex to index a
%particle. 
Set the weights $\{w^{(i)}(1|0)\}_{i=1}^M =1/M$ and form the particle
approximation of the predictive density
$p(\xx(1|0)|\ss(0))$ using \eqref{eq:partapp}. Set the design
parameter $\Delta t$, $\lambda$ and $\epsilon$. When
  $\yy(1)$ becomes available,  set $t=1$ and proceed to  step 1.
\subsection{Step 1: Propagate the Predictive Distribution
  One Step} In this step we propagate the predictive distribution $p(\xx(t|t-1)|\{\yy(s)\}_{s=1}^{t-1}, \ss(t-1))$ one time
step forward  and compute
$p(\xx(t+1|t)|\{\yy(s)\}_{s=1}^{t},\ss(t-1))$. Note that $\ss(t-1)$ is
kept fixed and we will deal with the update of $\ss$ and how that
changes the predictive distribution in step 2, 3 and 4, explained in
subsections C, D, and E respectively. 
%We will use  a particle approach and represent
%distributions using weighted sums of Dirac delta functions centered at particles.
%We  let $\xx^{(i)}(t+1|t)$  represents the $i$th particle at time $t+1$ using
%measurements up to time $t$. 
The  propagation of the predictive
distribution is done in three steps: 
\subsubsection{Measurement update}
%\begin{enumerate}
%\item[a)] {Measurement update: }
As $\yy(t)$ is made available, update the weights $\{w^{(i)}(t|t-1)\}_{i=1}^M $ according to
\begin{equation}
w^{(i)} (t) = \frac{ p(\yy(t)|\xx^{(i)} (t|t-1), \ss(t-1)) w^{(i)} (t|t-1)}{ \sum_{j=1}^M p( \yy(t)| \xx^{(j)} (t|t-1),\ss(t-1))
  w^{(j)} (t|t-1)},
\end{equation}
for  $i=1,\dots,M$. Since we assumed Gaussian additive measurement noise, 
\begin{equation}
p(\yy(t)|\xx^{(i)} (t|t-1),\ss(t-1)) = \N (\yy(t); h(\xx^{(i)} (t|t-1)), \RR) ,
\end{equation}
for $ i=1,\dots,M.$
%\item[b)] { Resample: } 
\subsubsection{Resample}
Resample using \eg \textit{sampling importance resampling} (SIR) \cite{gordon:93}.
%\item[c)] { Time update: }
\subsubsection{Time update}
Let  $ \xx^{(i)} (t+1|t) \sim p(\xx(t+1)|\xx^{(i)}(t), \ss(t-1)),
\,i=1,\dots,M.$ This can be done by sampling $\zz^{(i)} \sim
p_{\vv},\,i=1,\dots,M,$ and then let 
\begin{equation}  \xx^{(i)} (t+1|t) = diag(\ss(t-1)) \left
    (g(\xx^{(i)}(t),t)+\zz^{(i) }\right),\end{equation} for
$i=1,\dots,M$. Set
\begin{equation}
w^{(i)}(t+1|t) = w^{(i)}(t),\quad i=1,\dots,M. 
%\frac{p( \xx^{(i)} (t+1|t)  | \xx^{(i)} (t)   }{ q( \xx^{(i)}(t+1)
%  | x^i (t) ,  \{
%y(s)\}_{s=1}^t)}
\end{equation}

%Note that $p(\xx(t+1)|\xx^{(i)}(t))$ is given by
%\eqref{eq:dyneq}. 
%That is, if let $p_{\vv}$ be the pfd for $\vv$, we
%can write
%\begin{equation}
%p(\xx(t+1)|\xx^{(i)}(t)) =p_{\vv} ( xx(t+1);  \xx^{(i)}(t) )
%\end{equation}
%\end{enumerate}

% At time $t$, a particle filter (PF) is deployed to approximate the
% distribution over the  nonzero states $p(\xx_t|I(t),\yy(t))$. The model in the PF is
% \begin{subequations}\label{eq:dyn}
% \begin{align}
% \xx(t+1) = &diag(I(t)) \left (g(\xx(t),t)+v(t) \right)\\
% \yy(t)=&h(\xx(t) ,t) +w(t),
% \end{align}\end{subequations}
% and the support encoded in $\ss(t)$ is assumed to be equal to  $\ss(t-1)$
% Note that the particle filter does not run on the full state
% dimension but only the elements of $\xx$ that is in the support.  

%In general higher state dimension than output dimension.  Which make
%filtering difficult. 

\subsection{Step 2: Add Elements to the Support}
At each time, the likelihood for the given measurement given a
particle approximation of the expected value of $\xx(t)$ is 
  computed. We
do this by first computing an approximation of the expected value of
$\xx(t)$ by 
\begin{equation}\label{eq:mean}
\hat \xx(t) = \sum_{i=1}^M w^{(i)}(t) \xx^{(i)} (t).
\end{equation}
We then evaluate $\N (\yy(t) ; h(\hat \xx(t)),\RR)$ to get the
likelihood of $\yy(t)$ given $\hat \xx(t)$.  If
the likelihood is below some threshold, it is likely that an element
not in the support has become nonzero. The most natural thing to do would then be to seek the $\xx$,  that additional to the
elements indicated by $\ss(t-1)$ has one extra nonzero element, and that maximizes the likelihood $\N (\yy(t) ;
h( \xx),\RR)$. This problem can be shown given by: 
\begin{equation}\label{eq:qprob}
\begin{aligned}
\min_{\xx} & \quad\|\yy(t) -h( \xx)\|_{\RR} \\ \subjto &\quad
\|(\I-diag(\ss(t-1)) )\xx\|_0=1
\end{aligned}\end{equation}
This problem is unfortunately nonconvex. Inspired by the novel developments in NLCS and QBP, introduce \begin{equation}\X=\begin{bmatrix} 1 \\
  \xx \end{bmatrix}\begin{bmatrix} 1 & \xx \end{bmatrix}, \quad \PP= (\I -diag(\begin{bmatrix} 1 & \ss(t-1) \end{bmatrix} ) ).
\end{equation}
Problem \eqref{eq:qprob} can now be shown equivalent to 
\begin{equation} \label{eq:qprob2}\begin{aligned}
\min_{\X\succeq 0} & \quad\|\yy(t) -H(\X) \|_{\RR} \\ \subjto &\quad
 rank(\X)=1, \X(1,1)=1\\
& \quad  \|\PP
\X \PP\|_0=1,
\end{aligned}\end{equation}
if we define 
\begin{equation}
\Phi_i=\begin{bmatrix} a_i & \bb^\T_i /2 \\ \bb_i/2 &
  \Q_i \end{bmatrix}
\end{equation}
and the $i$th element of $H:\Re^{(n+1) \times (n+1)} \rightarrow
\Re^N$ as
\begin{equation}\big(
H(\X) \big)(i)=\Tr(\Phi_i\X).
\end{equation}
The rank and the zero-norm constraint make this problem nonconvex.   If we relax  the
rank constraint and instead minimizes a convex surrogate for the rank, we
obtain the optimization problem:
\begin{equation}\label{eq:qprob3}
\begin{aligned}
\min_{\X\succeq 0} & \quad\|\X\|_* + \lambda \|\yy(t) -H(\X) \|_{\RR} \\ \subjto &\quad
  \X(1,1)=1,\:
 \|\PP
\X \PP\|_0=1.
\end{aligned}\end{equation}
 $\|\cdot\|_*$ is here used to denote the nuclear norm, a well known
 convex surrogate for the rank of a matrix. Since $\X$ is constrained
 to be a positive semidefinite matrix, the nuclear norm is the same as
 the trace of $\X$.  $\lambda$ decides the tradeoff between the
 nuclear norm of $\X$ and the term $ \|\yy(t) -H(\X))
 \|_{\RR}$. If the solution $\X$ to \eqref{eq:qprob3} is not rank 1,
 $\lambda$ needs to be decreased. In general, the solution is rather
 insensitive to variations in $\lambda $ as long as a small enough
 $\lambda$ is chosen. The zero norm is now the only thing that makes
 this problem nonconvex. However, even though this problem is
 nonconvex, it is computationally tractable. Define the $(n+1)\times
 (n+1)$-dimensional matrix $\I_j $ as an identity matrix but with the $j+1$th
 diagonal element set to zero, \begin{equation}\I_j = diag( [1\cdots
   1\,0\,1\,\cdots 1] ) ). 
\end{equation} 
We now constraint all elements except one of $\PP \X \PP$ to be zero, say
that the $j$th element is allowed to be nonzero, and
solve the convex semi-definite program (SDP)
\begin{equation}\label{eq:qprob4}\begin{aligned} 
c(j)=\min_{\X\succeq 0} & \quad\|\X\|_* + \lambda \|\yy(t) -H(\X) \|_{\RR} \\ \subjto &\quad
  \X(1,1)=1,\:
 \I_j \PP_j
\X \PP_j=0.
\end{aligned}\end{equation}
repeatedly for $j=1,\dots,n.$ The $\X$ associated with the 
smallest $c(j) $ and a $j$ not already in the support, solves
\eqref{eq:qprob3}. Set $\ss(t-1)$  to reflect the support of
$\X(2:n+1,1)$ (the second to the last element of the first column of $\X$).

\subsection{Step 3: Remove Elements from the Support}
An index is removed from the support if the element associated with it
has an approximate
expectation value that for $\Delta t$ consecutive time steps  stays within some
epsilon bound of zero.  $\hat \xx(t)$ computed in \eqref{eq:mean} is used as an approximate
expectation value for $\xx(t)$.

\subsection{Step 4: Update the Predictive Distribution}
If the
  support has changed, set $\ss(t-\Delta t-1)=\ss(t-1)$. Compute the
  particle representation of the predictive density at time $t-\Delta t$,
  $p(\xx(t-\Delta t|t-\Delta t-1 )|\{\yy(s)\}_{s=1}^{t-\Delta t-1},
  \ss(t-\Delta t-1)),$ by updating the previously used particle
  representation for the predictive density at time $t-\Delta t$ as
  follows:
\begin{itemize}
\item If an element has been added to the support, use the initial
  distribution $p_0$ to initialize the new nonzero element in the
  particles $\{ \xx ^{(i)} (t-\Delta t|t-\Delta t-1)\}_{i=1}^M$.
\item If an element has been removed, set the corresponding element in
  $\{ \xx ^{(i)} (t-\Delta t|t-\Delta t-1)\}_{i=1}^M$ to zero.
\end{itemize}
Set $t=t-\Delta t$, and return to step
  1. Otherwise, set $\ss(t-1)=\ss(t)$, $p(\xx(t+1|t)|\{\yy(s)\}_{s=1}^{t}, \ss(t))=p(\xx(t+1|t)|\{\yy(s)\}_{s=1}^{t}, \ss(t))$, $t=t+1$ and return to step 1.

\section{Numerical Evaluation}
\begin{figure}[h!]
  \centering\includegraphics[width=0.9\columnwidth]{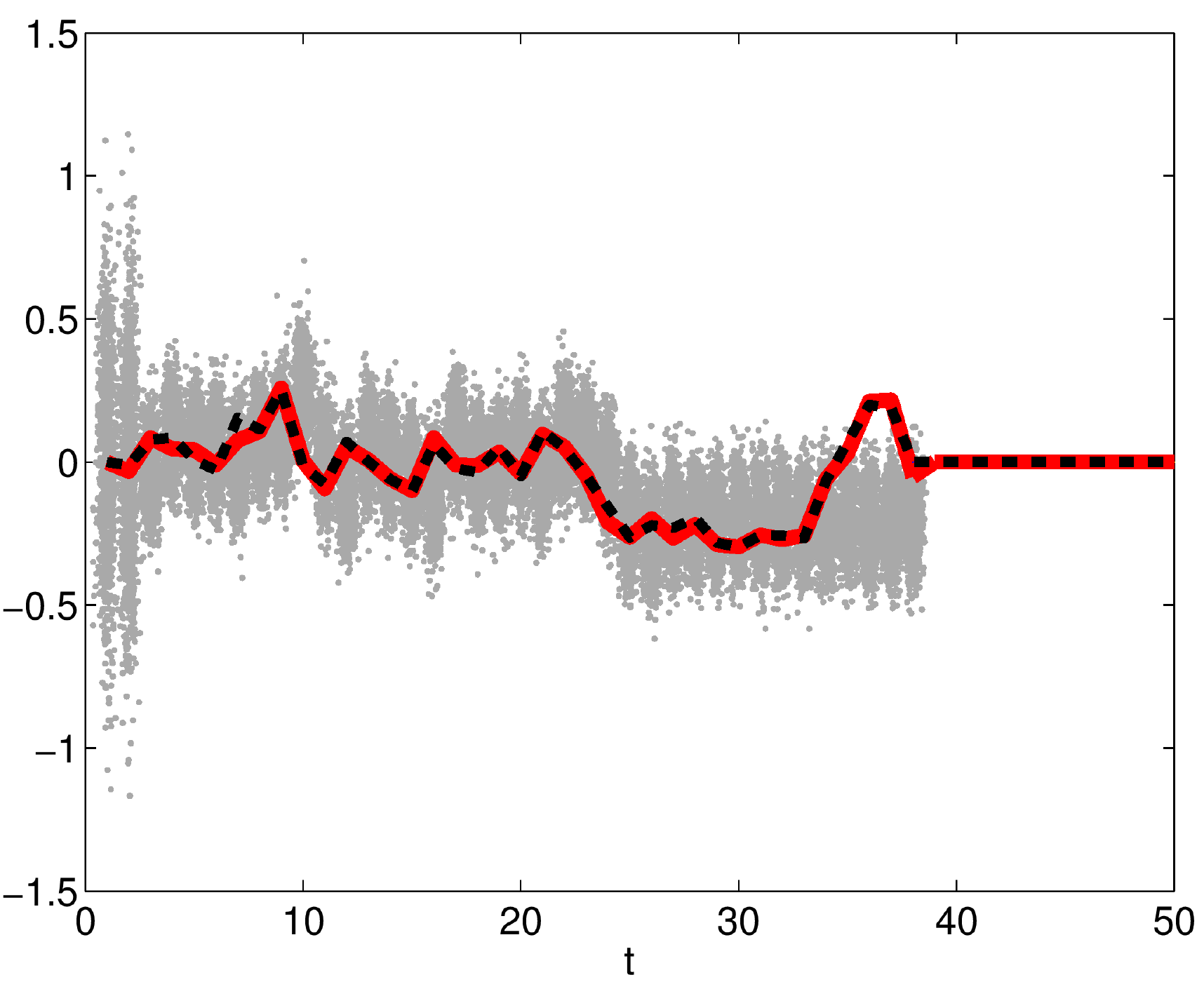}\caption{An
    illustratration of the particle density of NCPF, the approximate
    expectation value of NCPF and the true first element of
  the state.}\label{fig:first}\end{figure}

\begin{figure}[h!]
  \centering\includegraphics[width=0.9\columnwidth]{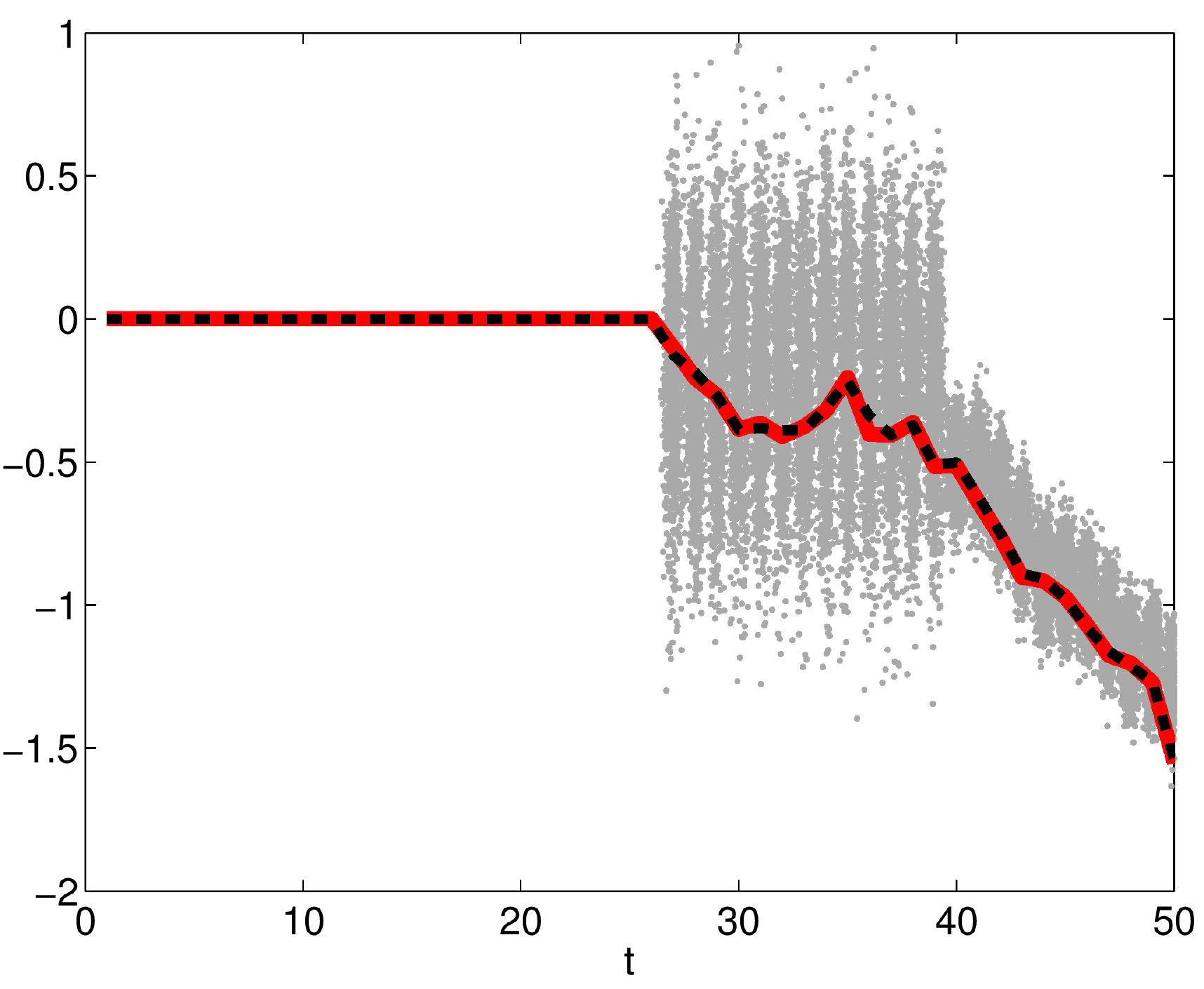}
\caption{An
    illustratration of the particle density of NCPF, the approximate
    expectation value of NCPF  and the true 27th element of
  the state.}\label{fig:sec}\end{figure}

In this example we  simulate data using the nonlinear dynamical model
\begin{subequations}\label{eq:dyn2}
\begin{align}\label{eq:dyneq2}
\xx(t+1) = &diag(\ss(t)) \left (\xx(t)+\vv(t) \right),\\
\yy(t)=&h( \xx(t))+\ww(t),\\
\ww(t) \sim & \N(0,0.01 \I), \quad \vv(t) \sim \N(0,0.01 \I).
%\\ h(\xx(t) ,t)=& a(t)+\bb(t)^\T \xx(t) +
%\xx^\T (t) \Q(t) \xx(t)
%\\  %\vv(t) \sim&\N(0,\QQ(t)),  \quad 
%\ww(t) \sim&
%\mathcal{U}(-\epsilon,\epsilon).
\end{align}\end{subequations}
$\ss(0)$ was set to $0$ and changed (either an element added or
removed) with probability 0.03. $a_i,\bb_i,\Q_i,i=1,\dots,N,$ were all generated
from a unitary Gaussian distribution. The state dimension was set to
$n=30$  and the measurement dimension to $20$. $M=10^6$, $\lambda=1$,
$\Delta t=3$ and the system was simulated  50 time steps. The whole
simulation, including  NCPF, took less than half a minute on a
standard laptop computer. Fig. \ref{fig:first} and \ref{fig:sec}
show, for a typical simulation, the
true state (dashed lines) and the estimated (solid lines) expectation
value of the state from the
NCPF. State 1 was nonzero between $t=0$ and $t=38$. At time $t=26$
element 27 in the state was added to the support. As seen in the
figure, the NCPF almost perfectly manage to track and detect the
correct state elements as nonzero.  Fig. \ref{fig:first} and
\ref{fig:sec} illustrate the particle densities for the first and 27th
element of the state as it propagates over time. 
% \begin{figure}[h!] \centering   \includegraphics[width=0.9\columnwidth]{fig/est2.pdf}
% \caption{The true state's nonzero components shown using black dashed
%   lines. The mean state of the NCPF shown using solid red and
%   green lines.}\label{fig:est}\end{figure}

We compare these results to that of a particle filter and to the
result obtained by ignoring the temporal dependence between states and
apply NLCS at each time instance.
\subsection{Particle Filter}
It is well known that particle filter can not handle problems of too
high dimension (the state dimension $n$ large). This is intuitively
understood from the fact that each particle
represent a hypotesis. As the dimension increase so does also the
different directions the noise can disturb the state etc. The number
of hypothesis needed can hence seen growing exponentially in the state
dimension. The number of particles needed to handle a low dimensional
system is therefore far fewer than for a high dimensional system.

\begin{figure}[h!] \centering\includegraphics[width=0.9\columnwidth]{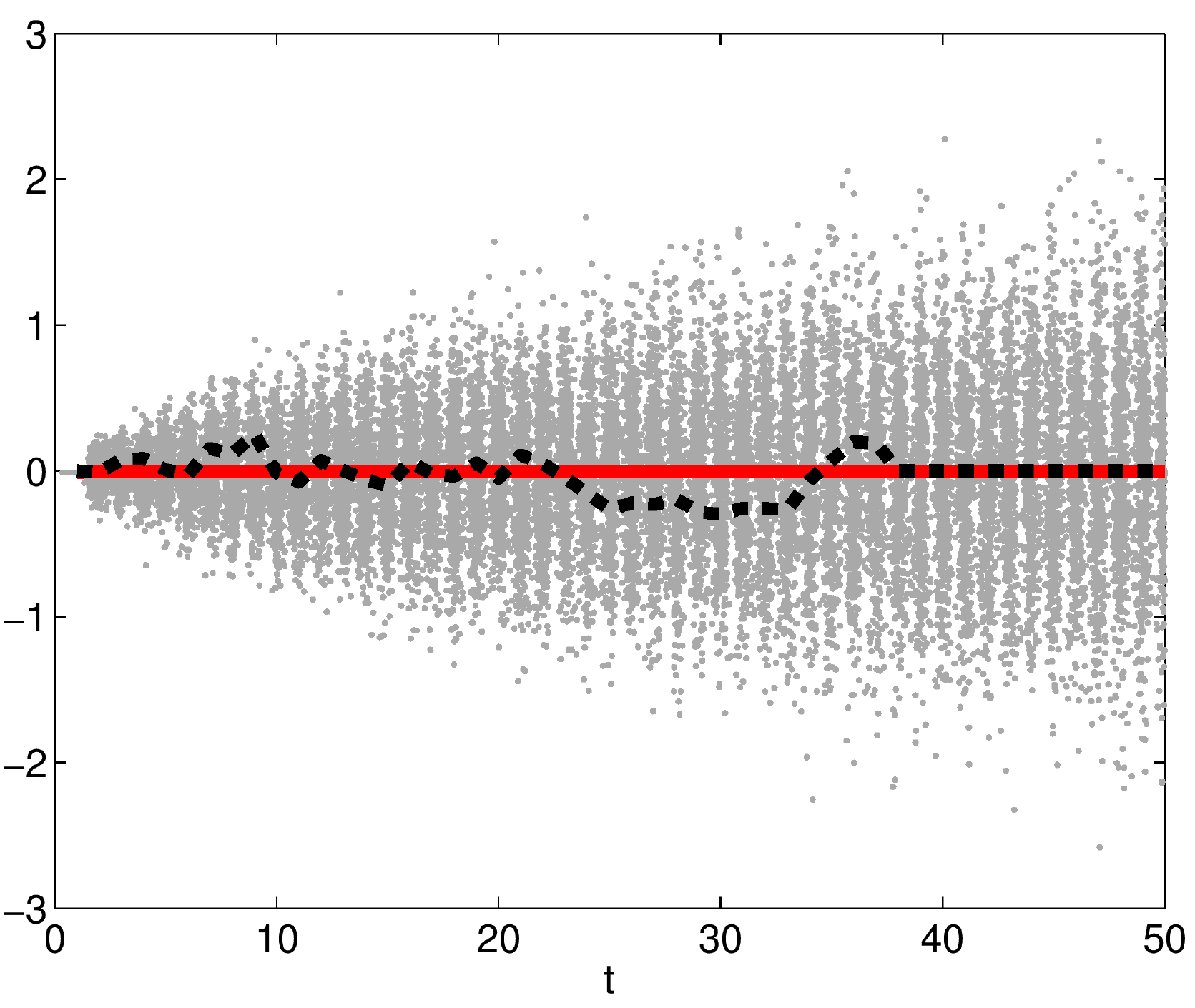}
\caption{The estimated expectation value of the first
element in the state for the particle filter, the true trajectory of the first element in the
state and a visualization of the particles' first
element.}\label{fig:pf}\end{figure}

Here we apply the particle filter with $10^6$ particles to estimate
the full 30 dimensional state. To be fair,
we initialized the particles to the true state (the support of the
true state at $t=1$ was given to the NCPF). The estimated expectation value for the first
element in the state, the true trajectory of the first element in the
state and a visualization of the particles are shown in  Fig.
\ref{fig:pf}. As seen, the particle filter totally fails to estimate
the first element of the state. Neither does it succeed in giving a
usable 
estimate for element 27. 

 \subsection{Nonlinear Compressive Sensing}
\begin{figure}[h!] \centering\includegraphics[width=0.9\columnwidth]{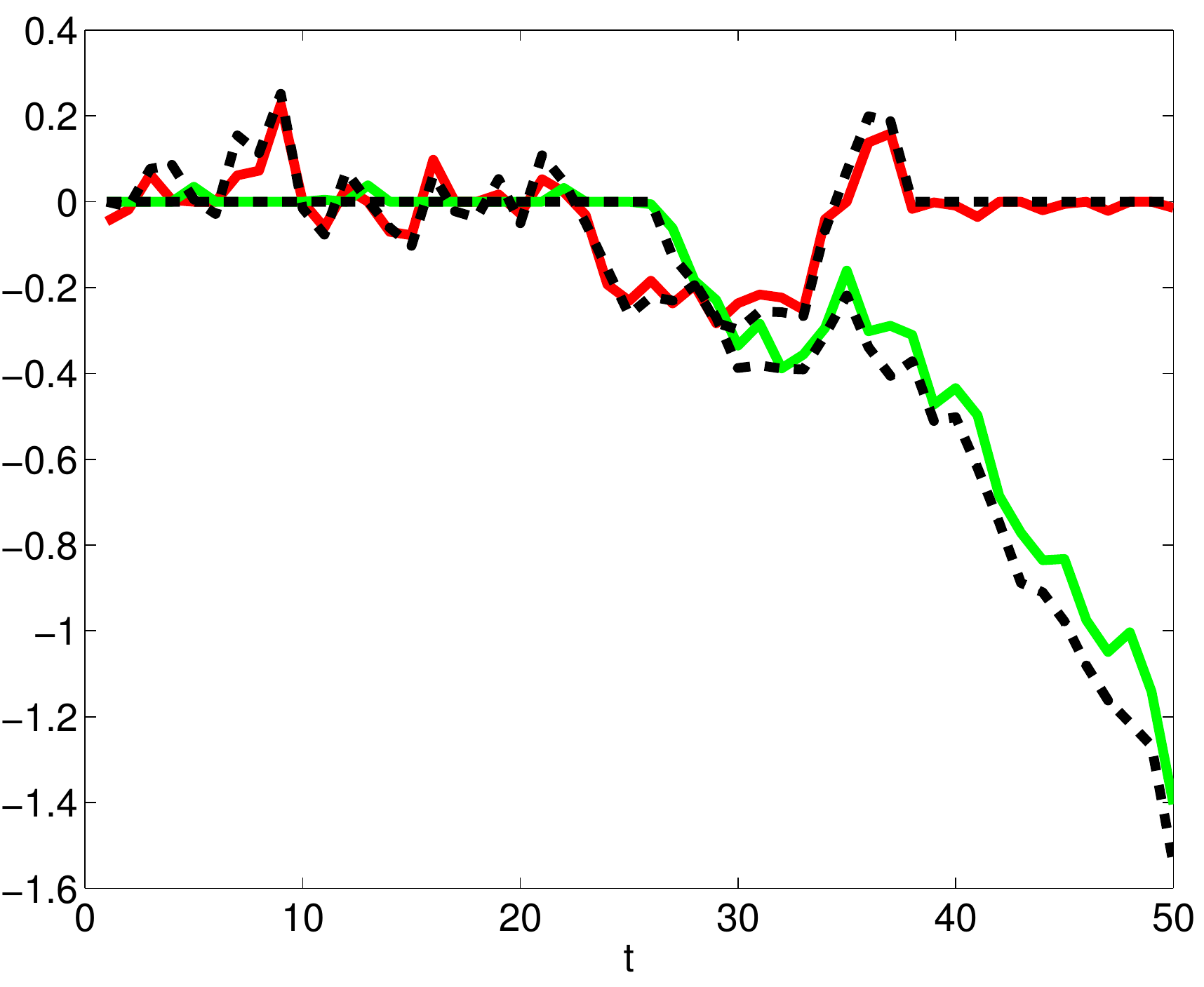}
\caption{Dashed line shows the true first and
27th state element and the solid line, NLCS estimates of the  first and
27th state element.}\label{fig:NLCS}\end{figure}

\begin{figure}[h!] \centering\includegraphics[width=0.9\columnwidth]{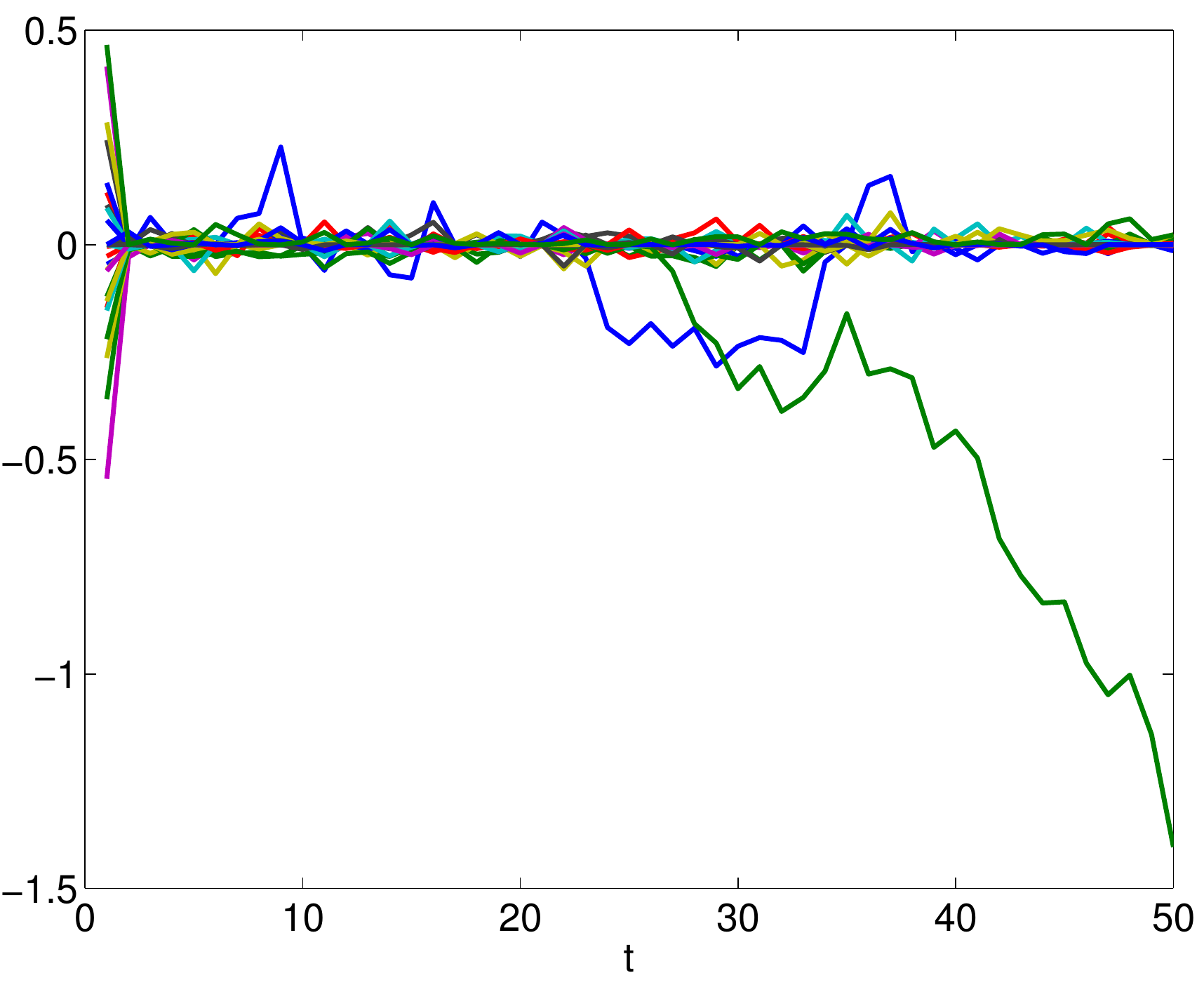}
\caption{An estiamte of  $\xx(t)$ using NLCS.}\label{fig:NLCS2}\end{figure}

NLCS can approximate the solution of  problems of the
form \begin{equation}\label{eq:NLCSest}\begin{aligned}\min_{\xx,\ww}
    \quad &\|\xx\|_0+\|\ww\|_\RR \\
\subjto \quad & \yy(t) =h(\xx)+\ww.\end{aligned}\end{equation}
If we neglect the time dependence, and as a new $\yy$ becomes
available use \eqref{eq:NLCSest} to find $\xx(t)$, we get the results
shown in Fig. \ref{fig:NLCS}. Dashed line shows the true first and
27th state element and the solid line, NLCS estimates of the  first and
27th state element. Fig. \ref{fig:NLCS2} shows the full state
estimates. As seen, NLCS gives comparable results with NCPF. However,
the computational complexity is not comparable and while the NCPF took
less than half a minute to run, the NLCS simulation took about 10 minutes. 

\subsection{Compressive Sensing and Some Remarks}
We also applied  ordinary linear compressive sensing to approximate the
state. The result was  worst than that shown above for the
NLCS. This is maybe  not
that strange since ordinary compressive sensing neglects the quadratic
part of the measurement equation.
We also combined ordinary compressive sensing with the particle
filter. This did not provide a satisfying result and compressive
sensing often failed to detect the correct state to be added.

Last we 
want to add that the weakest part in NCPF is not the mechanism for
adding or removing elements. The times we saw week performance, it was
because the particle filter diverged.

\section{Conclusion}
This paper proposes Nonlinear Compressive Particle Filtering (NCPF), a novel combination
of nonlinear compressive sensing and the particle filter. The
proposed framework is shown to work better than nonlinear compressive
sensing and particle filtering alone and offers an attractive solution
to nonlinear sparse filtering.

%%%%%%%%%%%%%%%%%%%%%%%%%%%%%%%%%%%%%%%%%%%%%%%%%%%%%%%%%%%%%%%%%%%%%%%%%%%%%%%%
% \section{ACKNOWLEDGMENTS}

% The authors gratefully acknowledge the contribution of National Research Organization and reviewers' comments.

%%%%%%%%%%%%%%%%%%%%%%%%%%%%%%%%%%%%%%%%%%%%%%%%%%%%%%%%%%%%%%%%%%%%%%%%%%%%%%%%

\bibliographystyle{plain}
\bibliography{refHO-1}
\end{document}